\pdfoutput=1
\RequirePackage{ifpdf}
\ifpdf 
\documentclass[pdftex]{sigma}
\else
\documentclass{sigma}
\fi

\newtheorem{prop}{Proposition}[section]
\newtheorem{lem}[prop]{Lemma}
\theoremstyle{definition}
\newtheorem{rem}[prop]{Remark}

\begin{document}

\newcommand{\inner}[1]{\left<#1\right>}

\newcommand{\ketbra}[2]{\left|#1\right>\!\left<#2\right|}

\allowdisplaybreaks

\renewcommand{\thefootnote}{$\star$}

\renewcommand{\PaperNumber}{062}

\FirstPageHeading

\ShortArticleName{Deformations of the Canonical Commutation Relations and Metric Structures}

\ArticleName{Deformations of the Canonical Commutation\\ Relations
and Metric Structures\footnote{This paper is a~contribution to the Special Issue on Deformations of Space-Time and its
Symmetries. The full collection is available at \href{http://www.emis.de/journals/SIGMA/space-time.html}
{http://www.emis.de/journals/SIGMA/space-time.html}}}

\Author{Francesco D'ANDREA~$^{a,\,c}$, Fedele LIZZI~$^{b,\,c,\,d}$ and Pierre MARTINETTI~$^{b,\,c}$}

\AuthorNameForHeading{F.~D'Andrea, F.~Lizzi and P.~Martinetti}

\Address{$^{a)}$~Dipartimento di Matematica e Applicazioni, Universit\`a di Napoli Federico II, Italy}
\EmailDD{\href{mailto:francesco.dandrea@unina.it}{francesco.dandrea@unina.it}}

\Address{$^{b)}$~Dipartimento di Fisica, Universit\`a di Napoli Federico II, Italy}
\EmailDD{\href{mailto:lizzi@na.infn.it}{lizzi@na.infn.it}, \href{mailto:martinetti.pierre@gmail.com}{martinetti.pierre@gmail.com}}

\Address{$^{c)}$~I.N.F.N.~-- Sezione di Napoli, Italy}

\Address{$^{d)}$ Departament de Estructura i Constituents de la Mat\`eria, Institut de Ci\'encies del Cosmos,\\
\hphantom{$^{d)}$}~Universitat de Barcelona, Spain}

\ArticleDates{Received March 02, 2014, in f\/inal form June 01, 2014; Published online June 10, 2014}

\Abstract{Using Connes distance formula in noncommutative geometry, it is possible to retrieve the Euclidean distance from the
canonical commutation relations of quantum mechanics.
In this note, we study modif\/ications of the distance induced by a~deformation of the position-momentum commutation
relations.
We f\/irst consider the deformation coming from a~cut-of\/f in momentum space, then the one obtained by replacing the usual
derivative on the real line with the~$h$- and~$q$-derivatives, respectively.
In these various examples, some points turn out to be at inf\/inite distance.
We then show (on both the real line and the circle) how to approximate points by extended distributions that remain at
f\/inite distance.
On the circle, this provides an explicit example of computation of the Wasserstein distance.}


\Classification{58B34; 46L87}

\renewcommand{\thefootnote}{\arabic{footnote}} \setcounter{footnote}{0}

\section{Introduction}

In this note we study metric structures induced by a~deformation of the canonical commutation relations of quantum
mechanics.
As explained by Connes and Marcolli in~\cite[Section~10.1]{Connes:2008kx}, it is precisely the lack of commutativity between
the coordinate and the translation operators which permits to retrieve the usual distance from a~purely
operator-algebraic setting.
Specif\/ically, by using the distance formula in noncommutative geometry~\cite{Con94} (see~\eqref{eq:one} below), one is
able to extract the Euclidean distance on $\mathbb{R}^d$, $d\geq 1$, from the canonical commutation relations (CCR)
\begin{subequations}
\label{eq:CCR}
\begin{gather}
[\boldsymbol{x}^\mu,\boldsymbol{x}^\nu]= 0,
\qquad
[\boldsymbol{p}_\mu, \boldsymbol{p}_\nu]=0,
\\
[\boldsymbol{x}^\mu, \boldsymbol{p}_\nu]=i\delta^\mu_\nu,
\label{eq:23}
\end{gather}
\end{subequations}
where $\mu,\nu=1,\ldots,d$, we take $\hbar =1$ and $\delta^\mu_\nu$ is the Kronecker symbol.
From this point of view, the CCR encode the metric structure of the \emph{classical} $\boldsymbol{x}$-space.
Let us stress that here by metric structure we mean a~distance, not a~metric tensor.

Several considerations, mostly inspired by quantum gravity, suggest that space (or spacetime) manifests its quantum
nature at very small scale.
One of the most elaborated theory of quantized spacetime is loop quantum gravity, where geometrical quantities such as
length, area and volume are encoded by operators with discrete spectra~\cite{Rovelli:1995fk}, and are in this sense
``quantized''.
However, the idea that sharp localization should become critical at very small length scales has been around for a~long
time~\cite{Bronstein} (see also~\cite{Piacitelli:2010uq} for an historical account).
Later on, it has been shown that gravity-induced uncertainty relations of Heisenberg type prevent the measurement of
position with arbitrary precision.
Models of quantum spacetime have been developed on this base, starting with a~noncommutativity of the coordinate
operators $\boldsymbol{x}^\mu$, $\boldsymbol{x}^\nu$ (see~\cite{Doplicher:1995hc} for the original paper,
and~\cite{Piacitelli:2010uq} for the recent developments).
String theory also indicates a~minimal uncertainty in space~\cite{AmatiCiafaloniVeneziano, GrossMende}, which can be
encoded in a~generalized uncertainty principle coming from a~deformation of the $\boldsymbol{x}$-$\boldsymbol{p}$
commutation relations~\cite{KempfManganoMann}\footnote{In these analyses, as well as in our study, not only the
algebraic commutation relations are important, but also their realization through operators on a~Hilbert space.
For instance, a~characteristic of~\cite{KempfManganoMann} is the use of symmetric operators with no selfadjoint
extensions.}. There are also models (e.g.~\cite{Amelino-Camelia:1998kx}) where both position-position and
position-momentum commutation relations are deformed.

In this paper, we focus on a~deformation of the $\boldsymbol{x}$-$\boldsymbol{p}$ relation only.
The coordinates~$\boldsymbol{x}^\mu$,~$\boldsymbol{x}^\nu$ still commute (as well as $\boldsymbol{p}^\mu$, $\boldsymbol{p}^\nu$).
From Gelfand--Naimark point of view, this means that topologically we are still dealing with a~classical space (the
algebra of coordinates is commutative).
However, since the distance formula of noncommutative geometry crucially depends on the commutators~\eqref{eq:23} (as
shown in~\eqref{eq:4}), a~modif\/ication of the CCR alters the metric.

One way to deform the CCR, inspired by regularization of renormalizable f\/ield theories~\cite{iraniani, Richter}, is
through the introduction of a~cut-of\/f on momenta.
This is discussed in Section~\ref{sec:2}.

\looseness=-1
In Section~\ref{sec:defoCCR} we study other modif\/ications of the CCR, focusing for simplicity on the one dimensional case.
We f\/irst consider the example where on the r.h.s.~of~\eqref{eq:23} there is an arbitrary function of $\boldsymbol x$,
which amounts to equipping the real line with an arbitrary Riemannian metric (Proposition~\ref{propnonflat}).
Then we consider a~commutator linear in $\boldsymbol{p}$ induced by a~deformation of the derivative
called~$h$-derivative
(see, e.g.,~\cite{KC02}).
We prove in Proposition~\ref{prop:2} that the distance between two points~$x$ and~$y$ is $|x-y|$ if $x-y\in h\mathbb{Z}$,
and is inf\/inite otherwise.
A~last example is the~$q$-derivative (see, e.g.,~\cite{Koo05}), corresponding to the~$q$-deformed relation $\boldsymbol
x\boldsymbol p -q\boldsymbol p \boldsymbol x=i$~\cite{FLW96}.
We prove in Proposition~\ref{prop:qder} that the distance is bounded from above by the ``French railway metric''.
We end the section with few considerations on the metric structure of the momentum space.

The picture which emerges from these various examples is that a~deformation of the CCR may alter the metric structure of
the line in a~fundamental way, such as points at inf\/inite distance (Proposition~\ref{prop:infinite}).
This invites to rethink the notion of point.
For instance, we show in Proposition~\ref{fatreal} how to approximate points of the real line by ``fat'' points
(rectangular distributions) which remain at a~f\/inite distance.
In Section~\ref{sec:6.2.1} we compute the distance for such fat points on the circle.
Proposition~\ref{prop:5.13} has an interest on its own (independent of noncommutative geometry), as an example of
optimal transport on the circle.

\section{Spectral metric geometry}
\label{sec:2}

This section contains some basics on the metric aspect of noncommutative geometry.
We show how, in the Euclidean case, Connes formula allows to retrieve the usual distance from the canonical commutation
relations of quantum mechanics.
Then, we recall how a~minimal distance on the position space may emerge from a~cut-of\/f in the momentum space.
Finally, we investigate the commutation relation coming from the cutof\/f.

\subsection{Euclidean distance from the canonical commutation relations}
Let~$M$ be an oriented Riemannian manifold without boundary, $\mathcal{A}:=C^\infty_0(M,\mathbb{R})$ the algebra of real
smooth functions vanishing at inf\/inity, $\mathcal{H}:=L^2(M,E)$ the Hilbert space of square integrable sections of
a~vector bundle $E\to M$, and represent $f\in \mathcal{A}$ on $\mathcal{H}$ by pointwise multiplication.
Let~$D$ be any self-adjoint (possibly unbounded) operator on $\mathcal{H}$ such that
\begin{itemize}\itemsep=0pt
\item[(i)] $[D,f]$ is a~bounded operator for any $f\in\mathcal{A}$.
\end{itemize}
Then for $x,y\in M$, the formula
\begin{gather}
\label{eq:one}
d_D(x,y):=\sup\bigl\{f(x)-f(y):\|[D,f]\|\leq 1\bigr\}
\end{gather}
def\/ines an extended metric on~$M$ (a distance, except that the value $+\infty$ is allowed).
Here the supremum is over all $f\in\mathcal{A}$ satisfying the side-inequality.

The distance~\eqref{eq:one} can be extended to arbitrary states $\varphi$, $\psi$ (positive linear functionals with
norm~$1$) of $C_0(M,\mathbb{R})$:
\begin{gather}
\label{eq:onebis}
d_D(\varphi,\psi):=\sup\bigl\{\varphi(f)-\psi(f):\|[D,f]\|\leq 1\bigr\}.
\end{gather}
Equation~\eqref{eq:one} is retrieved by considering pure states (extremal points of the convex space of states) which, by
Gelfand theorem, are the evaluation at points of~$M$.
Formula~\eqref{eq:onebis} is exactly the Kantorovich dual formula for the Wasserstein distance of order $1$
(see~\cite{DM09} and references therein).

When~$M$ is locally compact and complete and~$D$ is a~Dirac operator, it is well known that~\eqref{eq:one} coincides
with the geodesic distance of~$M$.
A~crucial property of (generalized) Dirac operators is that
\begin{itemize}\itemsep=0pt
\item[(ii)] $f(D+i)^{-1}$ is a~compact operator for all $f\in\mathcal{A}$.
\end{itemize}
A~triple $(\mathcal{A},\mathcal{H},D)$ satisfying (i) and (ii) is called a~\emph{spectral triple}, a~notion at the core
of Connes' noncommutative geometry~\cite{Con94}.
Accordingly, we call the extended metric~\eqref{eq:one} the \emph{spectral distance}.

On $\mathbb{R}^d$, $d\geq 1$, we interpret the Euclidean distance as the spectral distance def\/ined by phase-space
canonical commutation relations.
Indeed, the position and momentum operators $\boldsymbol{x}=\left\{\boldsymbol{x}^\mu\right\}$ and
$\boldsymbol{p}=\left\{\boldsymbol{p}_\mu\right\}$, $\mu=1,\ldots,d$, of~$d$-dimensional quantum mechanics are given, in
the space representation, by the operators
\begin{gather}
\boldsymbol{x}^\mu(\psi)(x)=x^\mu \psi(x),
\qquad
\boldsymbol{p}_\mu(\psi)(x)=-i\frac{\partial}{\partial x^\mu} \psi(x),
\label{eq:7}
\end{gather}
which are essentially selfadjoint on the domain $\mathcal{S}(\mathbb{R}^d)\subset L^2(\mathbb R^d)$ (the space of
Schwartz functions on $\mathbb{R}^d$).
Since the operators in~\eqref{eq:7} are endomorphisms of $\mathcal{S}(\mathbb{R}^d)$, their composition is well def\/ined
and the CCR~\eqref{eq:CCR} are satisf\/ied (for a~recent view on the CCR and its mathematical implications, e.g.,~dif\/ferent
choices of domains, see~\cite{Kadison:2014uq}).
Moreover, any $f\in C^\infty_0(\mathbb{R}^d)$ def\/ines a~bounded operator $f(\boldsymbol{x})$ on $L^2(\mathbb{R}^d)$
given by pointwise multiplication:
\begin{gather}
f(\boldsymbol{x})\psi(x):=f(x)\psi(x)
\qquad
\forall\,
\psi\in L^2\big(\mathbb{R}^d\big),
\quad
x\in\mathbb{R}^d.
\label{eq:9}
\end{gather}
The operator~\eqref{eq:9} maps the domain of self-adjointness of $\boldsymbol{p}_\mu$ into itself: the commutator
$[\boldsymbol{p}_\mu,f(\boldsymbol{x})]$ is then well-def\/ined, with closure the bounded operator $-i\partial_\mu
f(\boldsymbol{x})$.
If $D=\sum_\mu\gamma^\mu\boldsymbol{p}_\mu$ is the Dirac operator of $\mathbb{R}^d$, one easily checks that
(see, e.g.,~the proof of Proposition~2.1 of~\cite{DM09}):
\begin{gather*}
\|[D,f]\|=\sup_{x\in\mathbb{R}^d}\sqrt{\sum_\mu\left|\partial_\mu f(x)\right|^2}.
\end{gather*}
On the left hand side we have the operator norm on the Hilbert space of spinors, and on the right hand side the sup norm
of a~function of~$x$, i.e.~its operator norm as an operator on $L^2(\mathbb{R}^d)$.
Denoting the latter simply by $\|[\boldsymbol{p},f(\boldsymbol{x})]\|$ (it \emph{is} the operator norm of the commutator
for $d=1$), we can rewrite the spectral distance on $\mathbb{R}^d$ as
\begin{gather}
\label{eq:4}
d_{\boldsymbol{p}}(x,y):=\sup_{f\in
C^\infty_0(\mathbb{R}^d,\mathbb{R})}\big\{f(x)-f(y):\|[\boldsymbol{p},f(\boldsymbol{x})]\|\leq 1\big\},
\end{gather}
which of course for $\boldsymbol{p}$ as in~\eqref{eq:7}, satisfying the CCR, gives the Euclidean distance $|x-y|$.
In this sense, the metric structure of the position space is determined by the phase-space commutation relations.

Our aim in this paper is to study how the distance is modif\/ied if one takes position and momentum operators that satisfy
a~deformation of the CCR~\eqref{eq:23}.
In Section~\ref{cutoff} we consider a~regularized momentum operator $\boldsymbol{p}$; then in Section~\ref{sec:defoCCR}
we discuss the CCR associated to a~Riemannian metric on $\mathbb{R}$, and study various f\/inite-dif\/ference approximations
of the derivative.

\subsection{Minimal length by cutting-of\/f the geometry}
\label{cutoff}

If we introduce a~cut-of\/f in the `momentum space',
i.e.~we replace~$D$ in~\eqref{eq:one} by a~bounded operator~$D_\Lambda$
with norm~$\Lambda>0$, condition~(ii) is no longer satisf\/ied.
Nevertheless, condition (i) is trivially satisf\/ied and equation~\eqref{eq:one} still def\/ines an extended metric on~$M$.

Borrowing the terminology of~\cite{Con94}, since the length element ``$ds=D^{-1}$''
is no longer an inf\/initesimal
(i.e.~$f(D+i)^{-1}$ is no longer compact), it is reasonable to expect that points cannot be taken as close as we want
anymore, and a~minimum length will appear (from a~physical point of view, we cannot probe the space with a~resolution
better than $\Lambda^{-1}$).
Indeed, one has the following proposition.

\begin{prop}[\protect{\cite[Proposition~5.1]{DLM13}}]
\label{prop:1}
Let~$D$ be bounded with norm $\Lambda>0$.
Then for any $x\neq y$,
\begin{gather}
\label{eq:dxy}
d_D(x,y)\geq\Lambda^{-1},
\end{gather}
i.e.~the distance between two points cannot be smaller than the cut-off.
\end{prop}

Depending on the case, $d_{D_\Lambda}(x,y)$ might be inf\/inite for all $x\neq y$, and this would make the
inequality~\eqref{eq:dxy} not particularly interesting.
A~suf\/f\/icient condition to have points at inf\/inite distance is that the commutant of~$D$ in
$\mathcal{A}:=C^\infty_0(M,\mathbb{R})$ is non-trivial, i.e.~the set of $f\in\mathcal{A}$ such that $[D,f]=0$ contains
non-constant functions.

\begin{prop}
\label{lemma:0}
If there exists a~non-constant $f\in\mathcal{A}$ such that $[D,f]=0$, then there are $($at least$)$ two points at infinite
distance $d_D$.
\end{prop}

\begin{proof}
Since~$f$ is not constant, we can f\/ind $x$, $y$ such that $f(x)-f(y)\neq 0$.
Let $f_\lambda=\lambda f$.
One has $[D,f]=0$, hence $\|[D,f_\lambda]\|\leq 1$ and $d_D(x,y)\geq |f_\lambda(x)-f_\lambda(y)|=\lambda|f(x)-f(y)|$ for
all $\lambda>0$.
The inequality holds for all~$\lambda$, so $d_D(x,y)=\infty$.
\end{proof}

A non-trivial commutator however is not necessary to have inf\/inite distance, as it can be seen in the next proposition.

\begin{prop}[\protect{\cite[Proposition~5.4]{DLM13}}]
\label{prop:infinite}
Let~$D$ be a~finite-rank operator on $\mathcal{H}$.
For any $x\neq y$,
\begin{gather*}
d_D(x,y)=\infty.
\end{gather*}
\end{prop}

For instance, let $D=\ketbra{\psi_0}{\psi_0}$ be the rank $1$ projection in the direction of a~unit vector~$\psi_0$.
Then, $\|[D,f]\|$ is the uncertainty of~$f$ relative to the vector state~$\psi_0$ \cite[Lemma~5.3]{DLM13} (see also~\cite{Rie12}):
\begin{gather}
\label{eq:variance}
\|[D,f]\|^2=\inner{f\psi_0,f\psi_0}-\left|\inner{\psi_0,f\psi_0}\right|^2
\qquad
\forall\,
f\in\mathcal{A}.
\end{gather}
It follows from Cauchy--Schwarz inequality that~\eqref{eq:variance} is zero if and only if~$f$ is constant or $f\psi_0=0$.
In particular, if the support of $\psi_0$ is the whole~$M$ (e.g.,~$\psi_0$ is a~Gaussian for $M=\mathbb{R}^d$),
then~\eqref{eq:variance} vanishes only if~$f$ is constant, although $d_D$ is inf\/inite by
Proposition~\ref{prop:infinite}.

More generally, given a~spectral triple $(\mathcal{A}, \mathcal{H}, D)$, if the commutant of~$D$ in $\mathcal{A}$ is
$\mathbb{C} 1$ (unital case) or $\{0\}$ (non-unital case) we say that the seminorm $L_D=\|[D,\cdot]\|$ on $\mathcal{A}$
induced by the Dirac operator is \emph{Lipschitz}~\cite{Rie99}, or that~$D$ has \emph{trivial commutant} in
$\mathcal{A}$.
If $\mathcal{A}$ is unital with separable norm-closure, its representation is non-degenerate and~$D$ has a~compact
resolvent, then $d_D$ is f\/inite on the space of states of $\mathcal{A}$ as soon as $L_D$ is Lipschitz~\cite[Proposition~B.1]{Rennie:2006fk}.
Together with Proposition~\ref{lemma:0}, this shows in particular that on a~compact manifold the spectral distance $d_D$
def\/ined by an operator~$D$ with compact resolvent is f\/inite if and only if $L_D$ is Lipschitz.
In the light of Proposition~\ref{prop:infinite} one may wonder if an alternative characterization of the f\/initeness of
$d_D$ could be: \emph{assuming $L_D$ is Lipschitz, then $d_D$ is finite if and only~$D$ has compact resolvent}.
So far, we have no answer to this question.
Hints may come from Rief\/fel's characterization of the f\/initeness of the distance based on general properties of the
semi-norm $L_D$~\cite{Rieffel:1999wq}.

In the non-unital case the situation is more involved.
Instead of a~compact resolvent, one asks for condition (ii).
However the latter, together with the triviality of the commutant of~$D$, no longer guarantees that the distance is
f\/inite.
For instance, in the Moyal plane there are states (even pure) at inf\/inite distance from one
another~\cite{Cagnache:2009oe}, although the Dirac operator satisf\/ies (ii) and has a~trivial commutant in $\mathcal{A}$.
As well, on a~locally compact manifold with~$D$ the usual Dirac operator, there are states at inf\/inite distance from one
another (for instance a~state whose momentum of order~$1$ is inf\/inite is at inf\/inite distance from any point).
However two pure states (i.e.~two points) are at f\/inite distance as soon as~$M$ is locally compact and complete.
So one may wonder if the following holds: \emph{on a~locally compact and complete manifold, assuming~$D$ has trivial
commutant, the distance $d_D$ between points of~$M$ is finite if and only if~$D$ satisfies condition $(ii)$.}

\subsection{Commutator and edge ef\/fects}
We now discuss what type of position-momentum commutation relations are associated to the regularized Dirac operators
considered above.
For simplicity, we focus on the case $M=\mathbb{R}$.

A way to regularize the spectrum of the Dirac operator $D=-i\mathrm{d}/\mathrm{d} x$ of the real line is to replace it
with $D_\Lambda=P_\Lambda D$, where $P_\Lambda$ is the spectral projection of~$D$ in the interval $[-\Lambda,\Lambda]$.
In the momentum representation, the position and momentum operators $\widetilde{\boldsymbol{x}}$,
$\widetilde{\boldsymbol{p}}$ are the endormorphisms of $\mathcal{S}(\mathbb{R})$
\begin{gather*}
\widetilde{\boldsymbol{x}}(\psi)(p)=i\frac{\mathrm{d}}{\mathrm{d} p}\psi(p),
\qquad
\widetilde{\boldsymbol{p}}(\psi)(p)=p\psi(p),
\end{gather*}
obtained by conjugating $\boldsymbol{x}$ and $\boldsymbol{p}$ with the Fourier transform $\mathcal{F}$.
The truncated momentum operator is $\tilde{\boldsymbol{p}}_\Lambda=\mathbb{I}_\Lambda\tilde{\boldsymbol{p}}$, where
$\mathbb{I}_\Lambda:=\mathcal{F}P_\Lambda\mathcal{F}^{-1}$ is the operator of pointwise multiplication by the
characteristic function of the interval $[-\Lambda,\Lambda]$.
The range of $\widetilde{\boldsymbol{p}}_\Lambda$ is not in the domain of $\widetilde{\boldsymbol{x}}$ due to the
non-dif\/ferentiability of the characteristic function at $\pm \Lambda$, so that the commutator
$[\widetilde{\boldsymbol{x}},\tilde{\boldsymbol{p}}_\Lambda]$ is ill def\/ined.

To cure this, one can def\/ine $\widetilde{\boldsymbol{x}}$ on $\mathbb{I}_\Lambda\mathcal{S}(\mathbb{R})$ as
$\widetilde{\boldsymbol{x}} (\mathbb{I}_\Lambda\psi):= \mathbb{I}_\Lambda \widetilde{\boldsymbol{x}}\psi$ for all $\psi\in\mathcal{S}(\mathbb{R})$.
This is just the derivative of $\mathbb{I}_\Lambda\psi$, everywhere the latter is def\/ined, and is compactly supported hence $L^2$.
Then
\begin{gather*}
\inner{\varphi,\widetilde{\boldsymbol{x}}\widetilde{\boldsymbol{p}}_\Lambda \psi}
-\inner{\varphi,\widetilde{\boldsymbol{p}}_\Lambda\widetilde{\boldsymbol{x}}\psi}=
i\inner{\varphi,\mathbb{I}_\Lambda\psi}
\qquad
\forall\,
\varphi,\psi\in\mathcal{S}(\mathbb{R}),
\end{gather*}
which we interpret (in the weak sense) as the phase-space commutation relation:
\begin{gather}
\label{eq:xpLa}
[\widetilde{\boldsymbol{x}},\widetilde{\boldsymbol{p}}_\Lambda]=i\mathbb{I}_\Lambda.
\end{gather}

Another approach is by def\/ining the derivative of $\mathbb{I}_\Lambda\psi$, $\psi\in\mathcal{S}(\mathbb{R})$, in the
sense of distributions.
Let $\mathcal{S}'(\mathbb{R})$ be the space of tempered distributions and recall that any $L^2$-function~$f$ is the
integral kernel of a~tempered distribution $T_f$.
As usual, we think of~$\mathcal{S}(\mathbb{R})$ and $L^2(\mathbb{R})$ as subspaces of~$\mathcal{S}'(\mathbb{R})$ via the
monomorphism $f\mapsto T_f$.
Any $\xi\in\mathrm{End}\big(\mathcal{S}(\mathbb{R})\big)$ can be extended to an endomorphism of~$\mathcal{S}'(\mathbb{R})$ by duality.
In particular, $\widetilde{\boldsymbol{x}}(T)(f):=-T(\widetilde{\boldsymbol{x}}f)$
$\forall\,
T\in\mathcal{S}'(\mathbb{R})$, $f\in\mathcal{S}(\mathbb{R})$ is the derivative in the sense of distributions (besides
the~$i$ factor).
As well, any bounded operator $\xi$ on $L^2(\mathbb{R})$ can be extended to a~linear map
$L^2(\mathbb{R})\to\mathcal{S'}(\mathbb{R})$, $f\mapsto T_{\xi f}$; this in particular applies to~$\mathbb{I}_\Lambda$
and~$\widetilde{\boldsymbol{p}}_\Lambda$.
Integration by parts gives:
\begin{gather}
\label{eq:ccrdist}
\widetilde{\boldsymbol{x}}(T_{\widetilde{\boldsymbol{p}}_\Lambda f})(g)- T_{\widetilde{\boldsymbol{p}}_\Lambda
\widetilde{\boldsymbol{x}}f}(g)=-ipfg\big|_{-\Lambda}^{\Lambda} +iT_{\mathbb{I}_\Lambda f}(g)
\qquad
\forall\,
f,g\in\mathcal{S}(\mathbb{R}).
\end{gather}
Denoting by $\hat\delta_x:\mathcal{S}(\mathbb{R})\to\mathcal{S}'(\mathbb{R})$ the linear map
$\hat\delta_x(f)=f(x)\delta_x$, from~\eqref{eq:ccrdist} one obtains the identity of linear maps
$\mathcal{S}(\mathbb{R})\to\mathcal{S}'(\mathbb{R})$:
\begin{gather}
\label{eq:comxpL}
[\widetilde{\boldsymbol{x}},\widetilde{\boldsymbol{p}}_\Lambda]= i\mathbb{I}_\Lambda
-i\Lambda\big\{\hat\delta_\Lambda+\hat\delta_{-\Lambda}\big\}.
\end{gather}
In addition to~\eqref{eq:xpLa} which amounts to replacing the identity operator of the CCR by the cut-of\/f function
$\mathbb{I}_\Lambda$, the r.h.s.~of~\eqref{eq:comxpL} takes also into account the edge contributions due to the
non-dif\/ferentiability of the cut-of\/f.

Equations~\eqref{eq:xpLa} and~\eqref{eq:comxpL} are two ways to interpret the cut-of\/f on momenta in terms of a~modif\/ication
of the CCR, based on two possible extensions of the position operator.
Re\-gu\-larizing the momentum operator provides a~f\/irst example of modif\/ication of both the CCR and the spectral distance:
since $\|\boldsymbol{p}_\Lambda\|=\Lambda$, we know from Proposition~\ref{prop:1} that there is a~mi\-ni\-mum distance, that
is  $d_{\boldsymbol{p}_\Lambda}(x,y)\geq \Lambda^{-1}$ $\forall\, x\neq y$.
However, it is not clear how to compute the distance, or even f\/ind an upper bound.
It is an open problem to understand whether it is always inf\/inite, as in the f\/inite-rank case, or not.

\section{Spectral distance and phase-space commutation relations}
\label{sec:defoCCR}

In this section, we study other modif\/ications of the distance on $\mathbb{R}$ coming from deformations of the CCR: f\/irst
deformations associated with non-Euclidean metrics, then two approximations of the derivative on $\mathbb{R}$ by f\/inite
dif\/ference operators.
These commutation relations are not implemented by self-adjoint, nor even symmetric, operators.
For a~study of metric aspects this is not a~problem, since formula~\eqref{eq:4} makes sense also for non-selfadjoint
(and not even symmetric) momentum operators $\boldsymbol{p}$.

\subsection[Non-f\/lat metric: \protect{$[x, p]= G(x)$}]{Non-f\/lat metric: $\boldsymbol{[x, p]= G(x)}$}
\label{nonflat}

Given a~(Riemannian) metric tensor~$g$ on the real line, i.e.~a strictly positive smooth function, the associated
geodesic distance (for $x<y$) is
\begin{gather}
d_g(x,y):=\int_x^y\sqrt{g(t)}\mathrm{d}t.
\label{eq:14}
\end{gather}
If~$G$ is a~slowly increasing complex smooth function on $\mathbb{R}$ (hence, in the multiplier algebra of~$\mathcal{S}(\mathbb{R})$~\cite{Voi05}), the deformed momentum operator
\begin{gather*}
\boldsymbol{p}(f)(x):= -G(x)\partial_xf(x)
\end{gather*}
is a~well-def\/ined endomorphism of $\mathcal{S}(\mathbb{R})$ and satisf\/ies the commutation relation
\begin{gather*}
[\boldsymbol{x},\boldsymbol{p}]= G(\boldsymbol{x}).
\end{gather*}

\begin{prop}
\label{propnonflat}
Suppose~$G$ is never zero, $|G|^{-1}$ is smooth and the metric~\eqref{eq:14} with $g:=|G|^{-2}$ is complete.
Then the spectral distance $d_{\boldsymbol{p}}$ is the geodesic distance $d_g$.
\end{prop}
\begin{proof}
The norm inequality in~\eqref{eq:4} in the present case becomes
\begin{gather}
|f'(t)|\leq \sqrt{g(t)}
\qquad
\forall\,
t \in \mathbb R,
\label{eq:12}
\end{gather}
which integrated gives (for $x<y$):
\begin{gather}
|f(x)-f(y)|=\left|\int_x^yf'(t)\mathrm{d} t\right| \leq\int_x^y\sqrt{g(t)}\mathrm{d} t = d_g(x,y).
\label{eq:1}
\end{gather}
Hence $d_{\boldsymbol{p}}(x,y)\leq d_g(x,y)$.

Note that the function $f_y(t):=d_g(t,y)$ satisf\/ies~\eqref{eq:12} and $f_y(x)-f_y(y)=d_g(x,y)$, but it is not~$C_0^\infty$.
To prove that the inequality~\eqref{eq:1} is saturated, one uses a~sequence of $C^\infty_0$-approximations of~$f_y(t)$
(see the proof of~\cite[Proposition~2.1]{DM09} for details).
\end{proof}

\subsection[$h$-derivative: \protect{$[x,p]=i-hp$}]{$\boldsymbol{h}$-derivative: $\boldsymbol{[x,p]=i-hp}$}
\label{section-hderivative}
The~$h$-derivative of a~function~$f$ is def\/ined by~\cite{KC02}:
\begin{gather*}
D_h f(x)=\frac{f(x+h)-f(x)}{h},
\end{gather*}
where $h>0$ is a~real deformation parameter, and is a~bounded operator on $L^2(\mathbb{R})$.
It has norm $\Lambda=2h^{-1}$ and converges pointwise to the usual derivative
(if~$f$ is dif\/ferentiable) for $h\to 0$.
Let
\begin{gather}
\label{eq:notsym}
\boldsymbol{p}:=-iD_h=-ih^{-1}(U_h-1),
\end{gather}
with $U_t$ the translation operator
\begin{gather*}
(U_tf)(x):=f(x+t).
\end{gather*}
We obtain a~linear deformation of the CCR:
\begin{gather*}
[\boldsymbol{x},\boldsymbol{p}]=i-h\boldsymbol{p}.
\end{gather*}

\begin{rem}
The operator in equation~\eqref{eq:notsym} is not symmetric.
Alternatively we may consider the self-adjoint version $\boldsymbol{p}':=-\frac{i}{2h}(U_h-U_{-h})$.
However this would make the computation of the distance more dif\/f\/icult.
Notice also that this yields a~non-polynomial deformation of the CCR, for
$[\boldsymbol{x},\boldsymbol{p'}]=\frac{i}{2}(U_h+U_{-h})$ is not a~polynomial in $\boldsymbol{p'}$.
\end{rem}

The operator $\boldsymbol{p}$ is bounded but has no f\/inite rank.
From Proposition~\ref{prop:1} we know that $d_{D_h}(x,y)$ is greater than $\frac h2$, but we have no information on its f\/initeness.
We now show that two points are at f\/inite distance if and only if they are in the same orbit for the action of
$\mathbb{Z}$ generated by the translation $x\mapsto x+h$.
To this aim, we need the following lemma.

\begin{lem}
\label{lemma6.3}
For any $f\in C^\infty_0(\mathbb{R})$, we have $\|[D_h,f]\|=\|D_h f\|_\infty \leq \|f'\|_\infty$.
\end{lem}

\begin{proof}
From $D_h=h^{-1}(U_h-1)$, one deduces $[D_h,f]=(D_h f)U_h$.
Since $U_{h}$ is unitary, we get $\|[D_h,f]\|=\|D_h f\|_\infty= h^{-1}\sup\limits_{x\in\mathbb{R}}|f(x+h)-f(x)|$.
The inequality in the statement of the lemma then follows from $|f(x)-f(y)|\leq |x-y|\cdot\|f'\|_\infty$.
\end{proof}

\begin{prop}
\label{prop:2}
$d_{D_h}(x,y)=|x-y|$ if $x-y\in h\mathbb{Z}$, and is infinite otherwise.
\end{prop}
\begin{proof}
The distance is symmetric, so we assume $x>y$.
Take $x=y+nh$ with $n\geq 1$, and def\/ine $y_k:=y+kh$.
Lemma~\ref{lemma6.3} yields
\begin{gather*}
|f(x)-f(y)| \leq \sum_{k=0}^{n-1}|f(y_k)-f(y_k+h)|
\\
\phantom{|f(x)-f(y)|}{}
\leq n\sup_{t\in\mathbb{R}}|f(t)-f(t+h)|=nh \|D_hf\|_\infty= nh\|[D_h,f]\|.
\end{gather*}
Thus $d_{D_h}(x,y)\leq nh=|x-y|$.
Now, $\|[D_h, f]\|\leq \|f'\|_\infty$ implies the opposite inequality between the dual distance, that is $d_{D_h}(x,y)\geq |x-y|$.
This proves the f\/irst statement.

For $n>0$, consider the smooth function vanishing at inf\/inity:
\begin{gather}
\label{eq:fKx}
f_{n,x}(t)=ne^{-\frac{1}{n}\sqrt{1+(t-x)^2}}\sin^2 \tfrac{\pi(x-t)}{h}.
\end{gather}
Since the trigonometric part is~$h$-periodic, $D_h f_{n,x}(t)=n\sin^2 \tfrac{\pi(x-t)}{h}
D_he^{-\frac{1}{n}\sqrt{1+(t-x)^2}}$.
Ap\-plying Lemma~\ref{lemma6.3} to the exponential, whose Lipschitz norm is no greater than $1/n$, we get
\mbox{$\|[D_h,f_{n,x}]\|\leq 1$}.
This implies
\begin{gather*}
d_{D_h}(x,y)\geq |f_{n,x}(x)-f_{n,x}(y)| =ne^{-\frac{1}{n}\sqrt{1+(x-y)^2}}\sin^2 \tfrac{\pi(x-y)}{h}.
\end{gather*}
If $x-y\notin h\mathbb{Z}$, the lower bound is not zero, and goes to inf\/inity for $n\to\infty$.
\end{proof}

We wonder if there exists a~family of states convergent to pure states and whose distance is the Euclidean one.
Natural candidates are the rectangular distributions
\begin{gather}
\label{eq:28rect}
\delta_{x,\epsilon}(f)=\frac{1}{\epsilon}\int_{x-\epsilon/2}^{x+\epsilon/2}f(t)\mathrm{d} t,
\qquad
\epsilon>0,
\end{gather}
since the net $\{\delta_{x,\epsilon}\}_{\epsilon>0}$ is weakly convergent to $\delta_x$.
For a~f\/ixed~$\epsilon$, call $ \mathbb{R}_\epsilon=\{\delta_{x,\epsilon}\}_{x\in\mathbb{R}} $
\begin{prop}
\label{fatreal}
$(\mathbb{R}_\epsilon,d_{D_h})$ is a~metric space isometric to the Euclidean real line if\/f~$\epsilon$ is a~multiple
of~$h$:
\begin{subequations}
\begin{gather}
\label{eq:Rh}
\text{if}
\quad
\epsilon\in h\mathbb{Z}^+
\qquad
\text{then}
\qquad
d_{D_h}(\delta_{x,\epsilon},\delta_{y,\epsilon})=|x-y|
\qquad
\forall\,
x,y\in\mathbb{R},
\\
\label{eq:Rhbis}
\text{if}
\quad
\epsilon\notin h\mathbb{Z}^+
\qquad
\text{then}
\qquad
d_{D_h}(\delta_{x,\epsilon},\delta_{y,\epsilon})=
\begin{cases}
|x-y| &\text{if}\quad x-y\in h\mathbb{Z},
\\
\infty &\text{otherwise}.
\end{cases}
\end{gather}
\end{subequations}
\end{prop}
\begin{proof}
Assume that $x>y$.
From the inequality in Lemma~\ref{lemma6.3}, it follows
\begin{gather}
d_{D_h}(\delta_{x,\epsilon},\delta_{y,\epsilon})\geq d_{D}(\delta_{x,\epsilon},\delta_{y,\epsilon})=|x-y|,
\label{eq:45}
\end{gather}
where~$D$ is the usual derivative and the last equality comes, e.g.,~from~\cite[Proposition~3.2]{DM09}.

Let $\epsilon\in h\mathbb{Z}^+$ and~$n$ be the integer part of $(x-y)/\epsilon$, that is $x-y-\epsilon<n\epsilon\leq
x-y$.~Then
\begin{gather*}
\epsilon\delta_{x,\epsilon}(f) =\int_{n\epsilon-\epsilon/2}^{n\epsilon+\epsilon/2}f(t+y)\mathrm{d} t
+\int_{n\epsilon+\epsilon/2}^{x-y+\epsilon/2}f(t+y)\mathrm{d} t
-\int_{n\epsilon-\epsilon/2}^{x-y-\epsilon/2}f(t+y)\mathrm{d} t
\\
\phantom{\epsilon\delta_{x,\epsilon}(f)}
=\int_{-\epsilon/2}^{\epsilon/2}f(t+y+n\epsilon)
\mathrm{d} t
+\int_{n\epsilon-\epsilon/2}^{x-y-\epsilon/2} \left(f(t+y+\epsilon)-f(t+y)\right)\mathrm{d} t.
\end{gather*}
For any~$f$ with $\|[D_h,f]\|\leq 1$,
\begin{gather*}
\epsilon(\delta_{x,\epsilon}-\delta_{y,\epsilon})(f) =\int_{-\epsilon/2}^{\epsilon/2}\left(f(t+y+n\epsilon)-f(t+y)\right) \mathrm{d} t\\
\phantom{\epsilon(\delta_{x,\epsilon}-\delta_{y,\epsilon})(f)=}{}
+\int_{n\epsilon-\epsilon/2}^{x-y-\epsilon/2}\left(f(t+y+\epsilon)-f(t+y)\right)\mathrm{d} t
\\
\phantom{\epsilon(\delta_{x,\epsilon}-\delta_{y,\epsilon})(f)}{}
\leq n\epsilon\int_{-\epsilon/2}^{\epsilon/2}\mathrm{d}
t+\epsilon\int_{n\epsilon-\epsilon/2}^{x-y-\epsilon/2}\mathrm{d} t =n\epsilon^2+\epsilon(x-y-n\epsilon)=\epsilon(x-y),
\end{gather*}
where we use $f(t+y+n\epsilon)-f(t+y)\leq n\epsilon$, that follows from the norm condition.
Hence $d_{D_h}(\delta_{x,\epsilon},\delta_{y,\epsilon})~\leq |x-y|$. Equation~\eqref{eq:Rh} follows from~\eqref{eq:45}.

Similarly, for any~$\epsilon$ and any $x=y+nh$ with $n\geq 1$, one has
\begin{gather}
\label{eq:31ineq}
\delta_{x,\epsilon}(f)-\delta_{y,\epsilon}(f)=\frac{1}{\epsilon}\int_{-\epsilon/2}^{\epsilon/2}\left(f(t+x)-f(t+y)\right)
\mathrm{d} t
\leq \frac{nh}{\epsilon}\int_{-\epsilon/2}^{\epsilon/2}1\mathrm{d} t=nh=|x-y|.
\end{gather}
Hence the f\/irst equation in~\eqref{eq:Rhbis}.
The second one is obtained by considering the following function.
Let $f_{\infty,x}(t):=\sin^2 \tfrac{\pi (t-x)}{h}$, $e_n(t):=e^{-\frac{1}{n}\sqrt{1+t^2}}$ and $f_{K,n,x}(t)=K
e_n(t-x)f_{\infty,x}(t)$, with $n\in\mathbb{Z}^+$ and $K>0$.
Similarly to~\eqref{eq:fKx}, $\|[D_h,f_{K,n,x}]\|\leq 1$ for all $K<n$.
Since $e_n(t-y)$ is uniformly convergent to $1$ in the interval $[x-\epsilon/2,x+\epsilon/2]$, then
\begin{gather*}
\lim_{n\to\infty}\delta_{x,\epsilon}(f_{K,n,y})=K\delta_{x,\epsilon}(f_{\infty,y})=
\frac{K}{2\epsilon}\left[t-\frac{h}{2\pi}\sin\frac{2\pi t}{h}\right]^{x-y+\frac{\epsilon}{2}}_{x-y-\frac{\epsilon}{2}}.
\end{gather*}
Therefore, using some trigonometric identities:
\begin{gather*}
d_{D_h}(\delta_{x,\epsilon},\delta_{y,\epsilon})\geq
\lim_{n\to\infty}\big\{\delta_{x,\epsilon}(f_{{K,n},y})-\delta_{y,\epsilon}(f_{{K,n},y})\big\}=
\frac{Kh}{\pi\epsilon}\sin\frac{\pi\epsilon}{h}\sin^2\frac{\pi(x-y)}{h},
\end{gather*}
for all $K>0$.
If $\epsilon\notin h\mathbb{Z}^+$ and $x-y\notin h\mathbb{Z}$, the right hand side goes to inf\/inity for $K\to\infty$.
\end{proof}

Although the maximal resolution between points is of order $\|D_h\|^{-1}\sim h$, this result shows that the
Euclidean distance is recovered by considering non-pure states, like rectangular distributions with width $\epsilon=h$.

\begin{rem}
One may wonder what remains true for states that are more general than~\eqref{eq:28rect}.
Given a~positive integrable function~$\psi$ supported in the interval $[-\epsilon,\epsilon]$ and normalized to~$1$,
consider the corresponding family of states (for $x\in\mathbb{R}$):
\begin{gather*}
\Psi_{x,\epsilon}(f)=\int_{-\epsilon/2}^{\epsilon/2}\psi(t)f(t+x)\mathrm{d} t.
\end{gather*}
The inequality~\eqref{eq:31ineq} is still valid.
Thus $d_{D_h}(\Psi_{x,\epsilon},\Psi_{y,\epsilon})=|x-y|$ if $x-y\in h\mathbb{Z}$ (for any value of~$\epsilon$).
For $x-y\notin h\mathbb{Z}$, computing the distance is an open problem.
\end{rem}

\subsection[$q$-derivative: $xp-qpx=i$]{$\boldsymbol{q}$-derivative: $\boldsymbol{xp-qpx=i}$}

Another well-known ``approximation'' of the derivative is the~$q$-derivative~\cite{Koo05}
\begin{gather*}
D_qf(x)=\frac{f(x)-f(qx)}{(1-q)x}
\qquad
\forall\,
x\neq 0.
\end{gather*}
Here $0<q<1$ is a~deformation parameter.
The~$q$-derivative is extended by continuity at $x=0$: $D_qf(0)=f'(0)$ (provided the r.h.s.~is well-def\/ined).
Setting $\boldsymbol{p}=-iD_q$, one obtains the deformation of the CCR:
\begin{gather*}
\boldsymbol{x}\boldsymbol{p}-q\boldsymbol{p}\boldsymbol{x}=i.
\end{gather*}
This~$q$-deformed phase-space was studied in~\cite{FLW96}.
The momentum operator is not symmetric, and a~discussion on how to deform the conjugation operation in a~way that is
consistent with the commutation relations is in~\cite{FLW96}.

Because of the behavior at $x=0$, $D_q$ is not a~bounded operator, so one cannot expect a~minimum length.
We show that the distance between pure states is always f\/inite, and bounded by the ``French railway metric'':
$d_{\text{\textsc{sncf}}}(x,y):=|x|+|y|$ for any $x\neq y$, $d_{{\text{\textsc{sncf}}}}(x,x)=0$.
\begin{lem}
\label{eq:RL3}
For any $f\in C^\infty_0(\mathbb{R})$, we have $q^{\frac{1}{2}}\|[D_q,f]\|=\|D_qf\|_\infty\leq\|f'\|_\infty$.
\end{lem}

\begin{proof}
From the~$q$-analogue of the Leibniz rule, we get $ [D_q,f]\psi(x)=\psi(qx)D_qf(x) $ for all $x\in\mathbb{R}$ (including
$x=0$).
For $\rho>0$, let $T_\rho$ be the unitary operator def\/ined by $ T_\rho\psi(x)=\rho^{\frac{1}{2}}\psi(\rho x).
$ Then $[D_q,f]=q^{-\frac{1}{2}}(D_qf)T_q$, and
\begin{gather*}
[D_q,f]^*[D_q,f]=q^{-1}T_q^*|D_qf|^2T_q.
\end{gather*}
Hence $\|[D_q,f]\|^2=q^{-1}\|D_qf\|_\infty^2$.
The result follows from $|f(x)-f(y)|\leq |x-y|\cdot\|f'\|_\infty$.
\end{proof}

\begin{prop}
\label{prop:qder}
For any $x$, $y$,
\begin{gather}
\label{eq:6b}
q^{\frac{1}{2}}|x-y|\leq d_{D_q}(x,y)\leq q^{\frac{1}{2}}d_{\text{\textsc{sncf}}}(x,y).
\end{gather}
If further~$x$ and~$y$ are in the same orbit of $q^{\mathbb{Z}}$, then
\begin{gather}
\label{eq:6a}
d_{D_q}(x,y)=q^{\frac{1}{2}}|x-y|.
\end{gather}
\end{prop}

\begin{proof}
Assume $x>y$.
By Lemma~\ref{eq:RL3} we get $d_{D_q}(x,y)\geq q^{\frac{1}{2}}|x-y|$. If $y=q^nx$, with $n\geq 1$, then
\begin{gather*}
|f(x)-f(y)| \leq \sum_{k=0}^{n-1}|f(x_k)-f(x_{k+1})|
\nonumber
\\
\phantom{|f(x)-f(y)|}
\leq \|D_qf\|_\infty\sum_{k=0}^{n-1}(1-q)|x_k| =\|D_qf\|_\infty (1-q^n)|x|
=q^{\frac{1}{2}}\|[D_q,f]\||x-y|,
\end{gather*}
where $x_k:=q^kx$.
Thus $d_{D_q}(x,y)\leq q^{\frac{1}{2}}|x-y|$, which proves the f\/irst equation in~\eqref{eq:6b} and~\eqref{eq:6a}.

For $n\to\infty$, $y=q^nx\to 0$ and we get $|f(x)-f(0)|\leq q^{\frac{1}{2}}\|[D_q,f]\||x|$ for all $x\in\mathbb{R}$.
So, for any $x,y\in\mathbb{R}$:
\begin{gather*}
|f(x)-f(y)|\leq |f(x)-f(0)|+|f(0)-f(y)| \leq q^{\frac{1}{2}}\|[D_q,f]\|d_{\text{\textsc{sncf}}}(x,y),
\end{gather*}
which gives the upper bound in~\eqref{eq:6b}.
\end{proof}

Note that if $x$, $y$ are small, $d_{D_q}(x,y)$ can be as small as we want: as expected, there is no minimum length.
The geometrical reason why the distance is always f\/inite is that all the orbits of $q^{\mathbb{Z}}$ have a~common
accumulation point, given by $x=0$.

\subsection{Metrics on the momentum space}

We have shown by examples how a~modif\/ication of the CCR implies a~modif\/ication of the metric structure of the position
space.
To avoid confusion, let us stress that we are not dealing with a~``noncommutative space'' whose coordinates $x^\mu$, $x^\nu$ do not commute (the metric aspect of such spaces has been studied, e.g.,~in~\cite{Cagnache:2009oe,
Martinetti:2011fk, Martinetti:2011fkbis, Martinetti:2011fko}).
Here the point of view is the one of quantum mechanics, and the noncommutativity is between the coordinate operators of
the position space ($x$-space) and those of the momentum space ($p$-space).

Since we remain at a~formal level and are not considering any specif\/ic physical example, the distinction between the~$x$
and~$p$ spaces is mostly a~matter of convention.
One may as well use the CCR and its deformations to def\/ine an extended metric on the~$p$-space:
\begin{gather}
d_{\boldsymbol{x}}(q,p):=\sup_{f\in
C^\infty_0(\mathbb{R}^d,\mathbb{R})}\big\{f(q)-f(p):\|[\boldsymbol{x},f(\boldsymbol{p})]\|\leq 1\big\}.
\label{eq:pmetric}
\end{gather}
When the commutation relation is symmetric in $\boldsymbol{x}$-$\boldsymbol{p}$ (as for the undeformed CCR), one gets the same metric.
But this is not always the case: for the~$q$-derivative of Section~\ref{eq:RL3}, one also has to make the change $q\to q^{-1}$.
In the~$h$-derivative example $[\boldsymbol{x},\boldsymbol{p}]=i-h\boldsymbol{p}$ of Section~\ref{section-hderivative},
the~$p$-metric~\eqref{eq:pmetric} is the same as the~$x$-metric~\eqref{eq:4} coming from
$[\boldsymbol{x},\boldsymbol{p}]=-i+h\boldsymbol{x}$, which is a~example of non-f\/lat deformation of the~$x$-space as
studied in Section~\ref{nonflat}.

Another interesting example is the~$x$-metric for a~non-f\/lat deformation of the~$p$-space (equiva\-lently: the~$p$-metric
for a~non-f\/lat deformation of the~$x$-space):
\begin{gather*}
[\boldsymbol{x},\boldsymbol{p}]= G(\boldsymbol{p}),
\end{gather*}
where~$G$ is some well-behaved function.
For instance Kempf, Mangano and Mann~\cite{Kem97, KempfManganoMann} studied the quadratic deformation
\begin{gather*}
[\boldsymbol{x},\boldsymbol{p}]=i\big(1+\ell^2\boldsymbol{p}^2\big),
\end{gather*}
where $\ell>0$ has the dimension of a~length.
Formally one can obtain this commutation relation from the operator
\begin{gather*}
\boldsymbol{p}=-i\ell^{-1}\frac{U_\ell-U_{-\ell}}{U_\ell+U_{-\ell}},
\end{gather*}
and link the corresponding distance to the one of the~$h$-derivative $D_h$ for $h=2\ell$, noticing that
\begin{gather*}
D_{2\ell}f=i(U_{2\ell}+1)[\boldsymbol{p},f](U_{-2\ell}+1).
\end{gather*}
However, one has to be cautious with the domain of $(U_\ell+U_{-\ell})^{-1}$.

Notice that curved momentum spaces have recently been under investigation, since they are expected to be at the heart of
a~new view on relativity~\cite{A-CFK-GS}.

\section{Fat points and optimal transport on the circle}\label{sec:6.2.1}

In Section~\ref{section-hderivative} we show how to recover the Euclidean distance from a~deformed CCR using, instead of
points, non-pure states given by rectangular distributions.
In this section, we illustrate another interest for such ``fat points'' by computing the distance between rectangular distributions on the circle.
This gives an example where the distance between two translated states is neither the amplitude of translation (as in
the Euclidean space~\cite[Proposition~3.2]{DM09}), nor the geodesic distance of~$S^1$.
Furthermore, this result may have an interest on its own, independently of noncommutative geometry, since as far as we
know there are few examples of explicit computation of the Wasserstein distance on the circle (discrete distributions
have been recently studied in~\cite{J.-Rabin:2011fk}).

Let $\mathcal{A}=C^\infty(S^1, \mathbb R)$ and $D=-i \frac{\mathrm{d}}{\mathrm{d} x}$ the usual Dirac operator on the
circle (we think of functions on $S^1$ as $2\pi$-periodic functions on $\mathbb{R}$).
Consider the compactly supported distribution
\begin{gather*}
\delta_{x,\epsilon}(f):=\int_{-\epsilon}^{\epsilon}f(t+x)\mathrm{d}\mu_t,
\end{gather*}
where $0<\epsilon<\pi$ and $\mathrm{d}\mu_t$ is any distribution with support in $[-\epsilon,\epsilon]$ normalized to~$1$.
It is enough to compute the distance for $y=0$ and $0<x\leq\pi$ since the distance is translation invariant and symmetric~\cite[Lemma 5.9]{DLM13}.

\begin{figure}[th]\centering
\includegraphics[width=160mm]{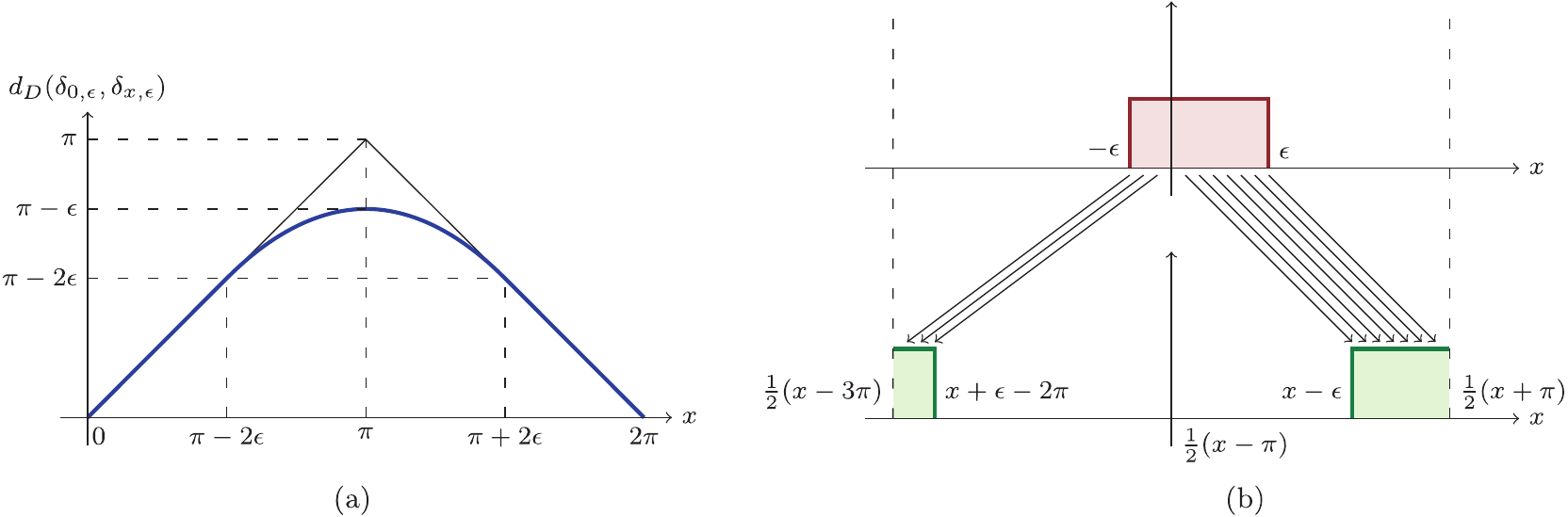}
\caption{The distance for rectangular distributions (a). The optimal transport map for rectangular distributions (b).}
  \label{fig1}
\end{figure}

\begin{prop}
\label{prop:5.13}
For $0< x\leq \pi-2\epsilon$ one has
\begin{gather}
d_{D}(\delta_{0,\epsilon},\delta_{x,\epsilon})=x.
\label{eq:48}
\end{gather}
For $\mathrm{d}\mu_t=\frac 1{2\epsilon}\chi_{[-\epsilon,\epsilon]}(t)\mathrm{d} t$ the rectangular distribution and
$\pi-2\epsilon~\leq~x\leq~\pi$, one has $($Fig.~{\rm \ref{fig1}(a))}
\begin{gather*}
d_{D}(\delta_{0,\epsilon},\delta_{x,\epsilon})= \tfrac{1}{4\epsilon}\big(-x^2+2\pi x-(\pi-2\epsilon)^2\big).
\end{gather*}
\end{prop}
\begin{proof}
Consider the $1$-Lipschitz function:
\begin{center}
\includegraphics[scale=0.85]{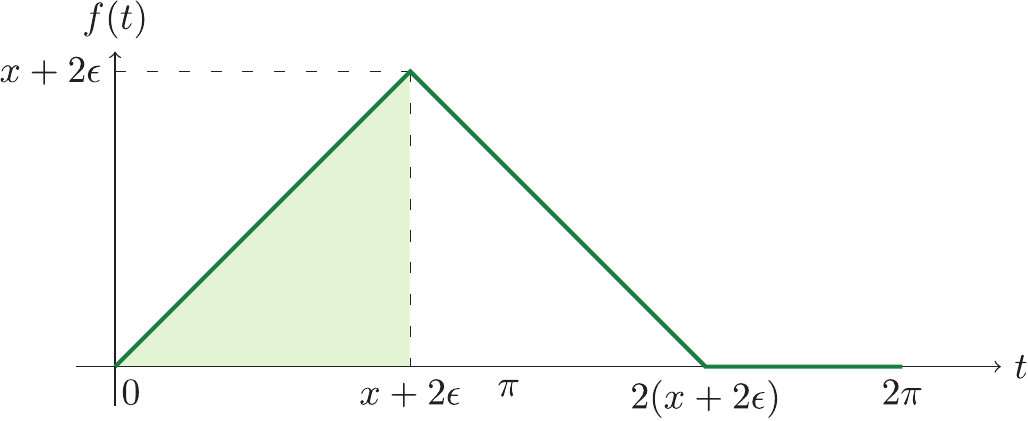}
\end{center}
With the replacement $f(x)\to f(x+\epsilon)$, we get for $0<x\leq\pi-2\epsilon$:
\begin{gather*}
d_{D}(\delta_{0,\epsilon},\delta_{x,\epsilon})\geq \int_{0}^{2\epsilon}\big\{f(t+x)-f(t)\big\}\mathrm{d}\mu_{t-\epsilon}
= \int_{0}^{2\epsilon}(t+x-t)\mathrm{d}\mu_{t-\epsilon} = x\int_{-\epsilon}^{\epsilon}\mathrm{d}\mu_t=x.
\end{gather*}
On the other hand for any periodic~$f$ with $\|[D,f]\|_\infty=\sup_t|f'(t)|\leq 1$ one has:
\begin{gather*}
|\delta_{0,\epsilon}-\delta_{x,\epsilon}|\leq\int_{-\epsilon}^\epsilon |f(t+x)-f(x)|\mathrm{d}\mu_t \leq\|f'\|_\infty\cdot x=x.
\end{gather*}
Hence the opposite inequality, that implies~\eqref{eq:48}.

For $\pi-2\epsilon \leq x\leq \pi$ and for rectangular distributions, the following function
\begin{center}
\includegraphics[scale=0.85]{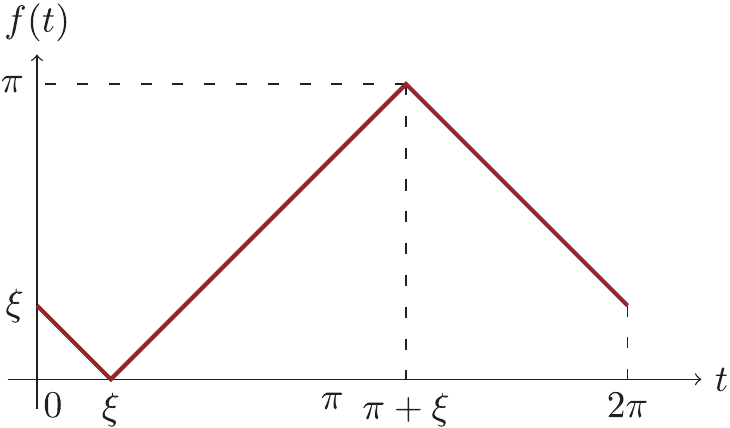}
\end{center}
yields
\begin{gather*}
2\epsilon d_{D}(\delta_{0,\epsilon},\delta_{x,\epsilon})\geq \int_{0}^{2\epsilon}\big\{f(t+x)-f(t)\big\}\mathrm{d}t
\\
\phantom{2\epsilon d_{D}(\delta_{0,\epsilon},\delta_{x,\epsilon})}
 =\int_{0}^{\xi}\big\{f(t+\pi+\xi)-f(t)\big\}\mathrm{d} t +\int_{\xi}^{2\epsilon}\big\{f(t+x -\xi)-f(t)\big\}\mathrm{d}x
\\
\phantom{2\epsilon d_{D}(\delta_{0,\epsilon},\delta_{x,\epsilon})}
 \geq \int_{0}^{\xi}\big\{(\pi-t)-(\xi-t)\big\}\mathrm{d}t+ \int_{\xi}^{2\epsilon}\big\{(t+x
-2\xi)-(t-\xi)\big\}\mathrm{d}t
\\
\phantom{2\epsilon d_{D}(\delta_{0,\epsilon},\delta_{x,\epsilon})}
 =\big\{(\pi-\xi)\xi+(x -\xi)(2\epsilon-\xi)\big\} =\tfrac{1}{2}\big({-}x^2+2\pi x -(\pi-2\epsilon)^2\big),
\end{gather*}
where $\xi:=\frac{1}{2}(x+2\epsilon-\pi)$.
On the other hand, for any $1$-Lipschitz function~$f$:
\begin{gather*}
f(t+\pi+\xi)-f(t)  \leq d_{\mathrm{geo}}(0,\pi+\xi)=2\pi-(\pi+\xi)=\pi-\xi,
\\  f(t+x -\xi)-f(t)  \leq d_{\mathrm{geo}}(0,x -\xi)=x -\xi,
\end{gather*}
where in the f\/irst equation we noticed that $\pi\leq\pi+\xi\leq 2\pi$.
Therefore
\begin{gather*}
d_{D}(\delta_{0,\epsilon},\delta_{x,\epsilon}) \leq \frac{1}{2\epsilon}\left(\int_{0}^{\xi}(\pi-\xi)\mathrm{d} t+
\int_{\xi}^{2\epsilon}(x-\xi)\mathrm{d} t\right)
\\
\phantom{d_{D}(\delta_{0,\epsilon},\delta_{x,\epsilon})}
=\frac{1}{2\epsilon}\big\{(\pi-\xi)\xi+(x-\xi)(2\epsilon-\xi)\big\} =\tfrac{1}{4\epsilon}\big(-x^2+2\pi
x-(\pi-2\epsilon)^2\big).
\end{gather*}
Hence the inequality is actually an equality.
\end{proof}

When $x+2\epsilon>\pi$ the distance is less than the geodesic one (Fig.~\ref{fig1}(a)) because one can optimize the
transport by moving part of the distribution to the left and part to the right along the circle (see Fig.~\ref{fig1}(b)).
As stressed in~\cite{Cabrelli:1995fk}, computing the Wasserstein distance on the circle amounts to cutting the circle at
a~well chosen point and then computing the same distance on the real line.
This cutting point is explicitly given in~\cite{Cabrelli:1995fk} for discrete distributions.
For rectangular distributions, one sees from the proof of Proposition~\ref{prop:5.13} that the cutting point has
coordinate $\frac{1}{2}(x-\pi)$ as in Fig.~\ref{fig1}(b).

The distance is also smoother than the Euclidean one (not at~$0$ though).
Interestingly the same phenomenon appears in a~totally dif\/ferent context (covariant Dirac operator on a~$U(n)$-bundle on
the circle~\cite{Martinetti:2006db,Martinetti:2008hl}).

\subsection*{Acknowledgements}

F.L.\ is partially supported by CUR Generalitat de Catalunya under project FPA2010-20807.
F.D.\ and F.L.\ were partially supported by UniNA and Compagnia di San Paolo under the grant ``Programma STAR 2013''.

\vspace{-2mm}

\pdfbookmark[1]{References}{ref}
\LastPageEnding

\end{document}